\documentclass[pre,aps,letterpaper,showpacs,floatfix]{revtex4-1}

\pdfoutput=1
\usepackage{amsmath}	
\usepackage{amsthm}
\usepackage{amsfonts}
\usepackage{amscd}
\usepackage{amssymb}

\usepackage{graphicx}	

\usepackage[utf8x]{inputenc} 	
\usepackage{type1cm}   			
\usepackage{mathtools} 			
\usepackage{sansmath}

\usepackage[euler]{textgreek}           

\usepackage{color}				

\begin{document}

%

\title{Structural Transitions in Fibers of Bent-Core Liquid Crystals from Field-Theory Monte Carlo Simulations}
\author{No\'e Atzin}
\affiliation{
Departamento de F\'{i}sica, Universidad Aut\'{o}noma  Metropolitana, Av. San Rafael Atlixco 186, Iztapalapa, Ciudad de M\'{e}xico, 09340, M\'{e}xico}
\author{Orlando Guzm\'an}
\email{ogl@xanum.uam.mx
}
\affiliation{
Departamento de F\'{i}sica, Universidad Aut\'{o}noma  Metropolitana, Av. San Rafael Atlixco 186, Iztapalapa, Ciudad de M\'{e}xico, 09340, M\'{e}xico}
\author{Juan J. de Pablo}
\affiliation{
Institute for Molecular Engineering, University of Chicago, Chicago, IL, USA}
\date{\today}

\begin{abstract}

Fibers of bent-core liquid crystals exhibit an internal structure consiting of a rolled smectic layer that can be used for optical waveguides. In this work, field-theoretic Monte Carlo simulations are used to analyze the internal configuration of such fibers as a function of the radial coordinate. We identify their equilibrium sates and we analyze the fully nonlinear model proposed by Bailey \textit{et al.} and revised by Perez-Ortiz \textit{et al}.

We find that, due to the non-differentiable character of such a model, the Euler-Lagrange equations are not able to find all equilibrium states. Our Monte Carlo procedure identifies both differentiable and non-differentiable equilibria and any first-order transitions between them. In all cases, the equilibrium states show inhomogenous configurations that display a boundary layer.

The methodology adopted here can by applied to other models of liquid crystals that have more degrees of freedom, including those with non-differentiable minima. The equilibrium structures presented here could be used as inputs for models of the transmission of light along the liquid crystal fibers.

\end{abstract}

%
%

\maketitle

%

\section{Introduction}

Liquid crystal phases have been studied extensively\cite{Onsager1949}, and have been found to have multiple technological applications. Their internal configuration can be easily manipulated with external fields or confining surfaces \cite{Lowe2012}, leading to uses as varied as in displays or sensors.  Experiments by Jákli \textit{et al.}\cite{Jakli2003} and, more recently, by Cheng \textit{et al}.\cite{Chen2013}, show that in contrast to calamitic mesogens, bent-core liquid crystals can form long, stable fibers.  These fibers, whose internal configuration consists of a spiral of smectic layers with well-aligned molecular dipoles, can be used as waveguides \cite{Fontana2009}.

Bailey \textit{et al.}\cite{Bailey2007} proposed a free energy model for such fibers as a function of the director orientation. By assuming a constant director orientation, these authors found the corresponding equilibrium states. Their model comprises three bulk contributions (Frank elasticity, layer-compression and electrostatic energies) and two surface terms (surface tension and divergence of molecular dipole orientation). P\'erez-Ortiz \textit{et al.}\cite{Perez-Ortiz2011} revisited this model using a variational approach and found that, in order to satisfy the boundary conditions inherent to such a system, there must exist a boundary layer with inhomogeneous director orientation. By assuming that the electrostatic contribution is small compared with the other energies, they linearized the Euler-Lagrange equations and solved them analytically, thereby showing that the boundary layer is about 100 nanometers thick.

Bauman \textit{et al.}\cite{Bauman2009} analyzed the stability of liquid-crystal fibers formed by bent-core mesogens using a generalization of a free energy model proposed by Bailey \textit{et al.} \cite{Bailey2007}: the fiber is described by concentric smectic layers, and the free energy is represented with a Landau expansion with orientational and strain elasticity, electric self interaction and dipolar divergence contributions, as well as surface tension. Assuming that the width of the smectic layers is small compared with the fiber radius, they proposed a criterion for the stability of the fibers: if the Frank energy is comparable to the energy for bending the smectic layer, then a circular fiber is stable.

To the best of our knowledge, however, past work has not considered the case when the electrostatic energy is of the same order of magnitude as other terms in the free energy, particularly when the director orientation is position dependent. This case is important because bent-core liquid crystals have relatively large spontaneous polarization $P_0$, as reported as early as 1991 by Niori et al.,\cite{Niori1996, Hird2005} and therefore large values of  electrostatic energy density.

As we show in Section \ref{freeEnergyModel}, the ratio of the free energy densities associated to spontaneous polarization and Frank elasticity is of the order of $P_0^2 L_B^2/(\epsilon_0 K)$, where $L_B$ is a characteristic length for changes in the director, $\epsilon_0$ is the permittivity of vacuum and $K$ is a Frank elastic constant. Assuming the values $P_0$= 50 nC cm$^{-2}$ reported by Niori et al.,\cite{Niori1996} $L_B$ = 100 nm reported for the boundary layer by Pérez-Ortiz et al.,\cite{Perez-Ortiz2011} and a typical value $K = 10^{-11}$ N,\cite{Bailey2007} we find that the electrostatic energy density associated with $P_0$ is about thirty times larger than the elastic energy density.  Even larger values of spontaneous polarization may be obtained using recently reported mesogens with polarized metallorganic complexes at the tip of their bent cores: Ohtani and coworkers reported a value $P_0$ = 1.08 $\mu$C cm$^{-2}$ for a bent-core liquid crystal synthetized with oxovanadium complexes.\cite{Ohtani2015}

In this work, we propose a field-theory Monte Carlo simulation where the degrees of freedom are provided by the director field. Using the fully nonlinear free energy model of {P\'erez-Ortiz} \textit{et al.}, we study the case when the electrostatic contribution cannot be neglected in comparison with the  elastic or the layer compression terms.
We search for the equilibrium states, and find that they have boundary layers of widths comparable to those predicted by {P\'erez-Ortiz} \textit{et al.}
Building on that finding, Monte Carlo simulations are also used to identify the equilibrium states that occur for configurations where the free energy model is non-differentiable. Such configurations are necessarily missed by the Euler-Lagrange equation formulation. We find first-order transitions between metastable and stable equilibria of differentiable and non-differentiable character.

Finally, using values for the material parameters reported previously \cite{Bailey2007,Perez-Ortiz2011}, our simulations are used to predict the radii of fibers in equilibrium in the same range as those observed experimentally. These radii are highly sensitive to the coefficient of the electrostatic energy, $c'$, in the sense that doubling the value of this parameter leads to an increment of one order of magnitude of the equilibrium radius.

\section{Free energy model}
\label{freeEnergyModel}

\begin{figure}[htb]
\begin{center}
\includegraphics[width=8cm,height=!]{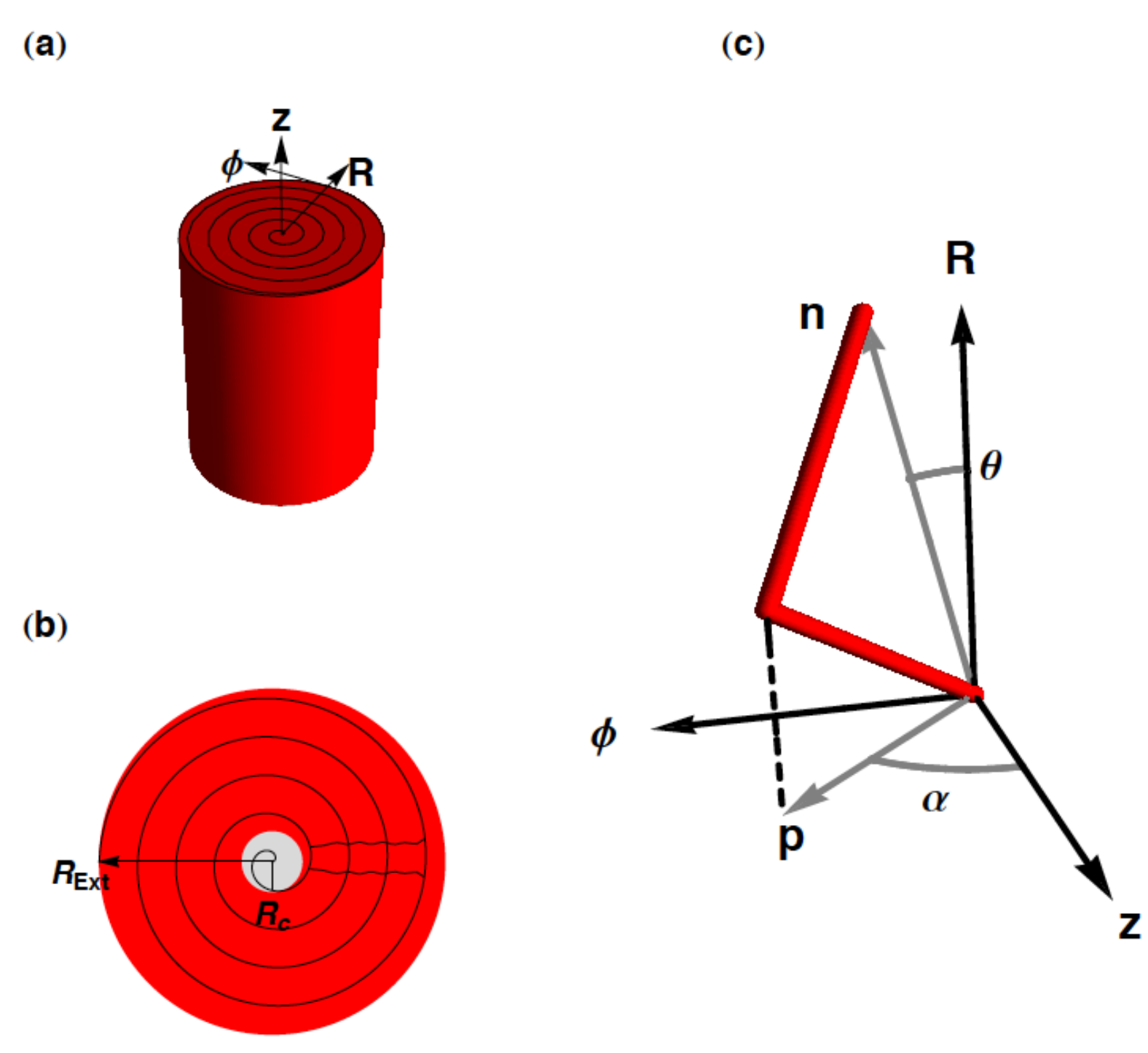}
\end{center}
\caption{ (a) We describe the structure of a liquid crystal fiber using a cylindrical coordinate system. (b) A cross section of the fiber shows the external and core radii and the spiral smectic layer. (c) At each position within the fiber, the orientation of mesogens is defined by the director $\pmb{n}$ and the polarization vector $\pmb{p}$  (parametrized by angles $\theta$ and $\alpha$, respectively).}
\label{Fig1}
\end{figure}

From the experiments of Chen \textit{et al.} \cite{Chen2013}, we take the structure of the LC fiber to be that of a rolled smectic layer. Thus, we model the fiber as a cylinder having an internal spiral arrangement (see Fig.~\ref{Fig1}~a). The fiber has an external radius $R_f$ and the topological defect at the center is taken to have a radius $R_c \ll R_f$ (see Fig~\ref{Fig1}~b).  By following a radial trajectory from the center to the surface of the fiber, we encounter a series of smectic layers. Making the approximation that the director orientation changes very slowly with the azimuthal coordinate ($\phi$) in comparison with the radial one ($r$), we take the director's orientation $\theta$ to be a function of $r$ alone, just as proposed by Perez-Ortiz \textit{et al.} \cite{Perez-Ortiz2011}.

Our model for the free energy is similar to that of Bailey \textit{et al.} \cite{Bailey2007} but without the assumption of a uniform orientation field. We describe our free energy model in terms of the orientation field of the liquid crystal through three orthonormal vectors: $\pmb{n}$ is the director, $\pmb{p}$ is the molecular dipole vector and $\pmb{m}= \pmb{n} \times \pmb{p}$. This orthonormal basis is defined by the following parameterization:
\begin{eqnarray}
\pmb{n} &=& \cos {\theta} \ \hat{\pmb{R}} \ + \ \sin{\theta} \ \hat{\pmb{\phi}}, \nonumber \\
\pmb{p} &=& \sin{\theta} \ \sin{\alpha} \ \hat{\pmb{R}} \  - \ \cos{\theta} \ \sin{\alpha} \ \hat{\pmb{\phi}} \ + \ \cos{\alpha} \ \hat{\pmb{z}
}, \\
\pmb{m} &=& \sin{\theta} \ \cos{\alpha} \ \hat{\pmb{R}}  \ - \ \cos{\theta} \ \cos{\alpha} \ \hat{\pmb{\phi}} \ - \ \sin{\alpha} \ \hat{\pmb{z}}, \nonumber
\end{eqnarray}
where $\hat{\pmb{R}}$, $\hat{\pmb{\phi}},$and $\hat{\pmb{z}}$ are the cylindrical-coordinates orthonormal basis. These vectors are illustrated in Fig.~\ref{Fig1}.

As mentioned above, we model the liquid crystal fiber in a field-theory framework, using the director $\theta(r)$. The fiber is described through the interplay of three bulk free energy contributions plus two surface terms:
\begin{equation}
\mathcal{F} = \int_\Omega  f_\text{N}  + f_\text{L} + f_\text{E} \ \text{d}V + \int_{\partial\Omega}  f_\text{S} + f_\text{D} \    \text{d}S,
\end{equation}
where the volumetric free-energy densities $f_\text{N}$, $f_\text{L}$ and $f_\text{E}$ correspond to orientational elasticity, layer compression elasticity, and dielectric contributions, respectively. The surface free-energy densities $f_\text{S}$ and $f_\text{D}$ model the effect of anisotropic surface tension and a contribution associated with inhomogeneities of polarization direction of the LC. These terms have been discussed by Bailey \textit{et al}. \cite{Bailey2007} and P\'{e}rez-Ortiz \textit{et al.} \cite{Perez-Ortiz2011} and only a brief account is included in what follows.

\subsection{Bulk free-energy densities}

Since we do not assume that the orientation of the LC is homogeneous inside the fiber, we have to consider the effect of distortions in the orientation. For this, we use the Frank-Oseen expression for liquid crystal elasticity, in terms of spatial derivatives of the director \cite{deGennes1993,Yokoyama1997},
\begin{equation}
f_\text{N}= {K_{11}\over 2} \left( \pmb{\nabla} \cdot \pmb{n} \right)^2 + {K_{22} \over 2} \left[  \left( \pmb{\nabla} \times \pmb{n} \right) \cdot \pmb{n}  \right]^2 + {K_{33}\over 2} \left[ \pmb{n} \times \left(  \pmb{\nabla} \times \pmb{n} \right)  \right]^2.
\end{equation}
These terms consider splay, twist and bend modes and the coefficients $K_{ii}$ are the respective elastic constants. Using the one-constant approximation, $K_{11}=K_{22}=K_{33}=K$, in cylindrical coordinates the Frank-Oseen density simplifies to
\begin{equation}
f_\text{N}= {K\over 2} \left[ {1 \over R^2} + \left( {\text{d} \theta \over \text{d} R  } \right)^2  \right].
\end{equation}
We follow Bailey \textit{et al.} and P\'{e}rez-Ortiz \textit{et al.} and study the bulk free-energy of the fiber by neglecting the contribution from the defect-core region \cite{Bailey2007}, which has a size of the order of the coherence length $\delta=\sqrt{K/B}$  (for our system, $\delta=10$~nm) \cite{Perez-Ortiz2011}.

\begin{figure}[ht]
\begin{center}
\includegraphics[width=8cm,height=!]{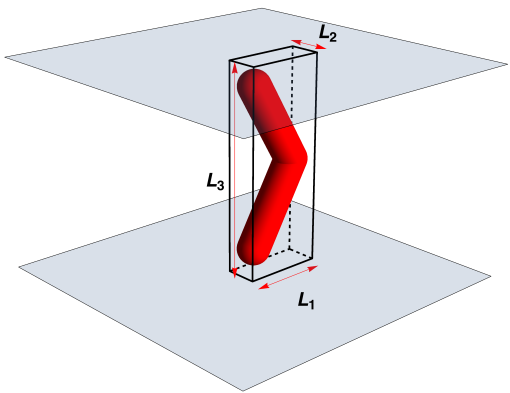}
\end{center}
\caption{The layer-compression free energy is determined by the orientation of the liquid crystal in the fiber. It has been parametrized by Bailey \textit{et al}.\cite{Bailey2007} in terms of the width ($L_1$), depth ($L_2$) and height ($L_3$)  of a box enclosing the bent-core mesogens.}
\label{Fig2}
\end{figure}

As the mesogens in a given layer change orientation, the layer width will also change. We model the layer-compression free energy density through the layer's strain ($\gamma$) produced when the width changes from its value $L_0$ in the flat-layer smectic to a value $L$ inside the fiber,
\begin{eqnarray}
f_\text{L} =  {B \gamma^2 \over 2} =  {B \over 2} \left( L- L_0 \over L_0 \right)^2,
\end{eqnarray}
where $B$ is the layer compression modulus. The orientation dependent width $L(\theta,\alpha)$ has been estimated by Bailey \textit{et al}. in terms of the local orthonormal vectors $\mathbf{n}, \mathbf{p}$ and $\mathbf{m}$ for the liquid crystal \cite{Bailey2007}:
\begin{eqnarray}
L\left(\theta, \alpha \right) = L_3 |\cos \theta | + L_2 | \cos \alpha \sin \theta |  + L_1 | \sin \alpha \sin \theta |.
\label{LayerWidth}
\end{eqnarray}
The constants $L_1$, $L_2$ and $L_3$ correspond to the width, depth and height of a rectangular box encasing a bent-core mesogen, as shown in Fig.~\ref{Fig2}.
For the flat-layer width, we introduce the flat-layer orientation angles $\Theta_0$ and $\alpha_0$ and substitute them in Eq.(\ref{LayerWidth}):
\begin{eqnarray}
L_0= L(\Theta_0, \alpha_0)= L_3 | \cos \Theta_0| + L_2 | \cos \alpha_0 \sin\Theta_0| + L_1 | \sin \alpha_0 \sin\Theta_0|.
\end{eqnarray}

When mesogens pack parallel to each other inside the layers, a spontaneous polarization $\pmb{P}_s$ is produced in addition to the  polarization proportional to the electric field. Then, the constitutive relation for the electric displacement is
\begin{eqnarray}
\pmb{D} = \pmb{P}_s + \epsilon_0 \; \mbox{\sf\straightepsilon} \cdot \pmb{E},
\end{eqnarray}
where
$ \mbox{\sf\straightepsilon} = \mbox{\straightepsilon}_1\; \pmb{n} \pmb{n} + \mbox{\straightepsilon}_2\; \pmb{m} \pmb{m} + \mbox{\straightepsilon}_3 \; \pmb{p} \pmb{p}$ is the dielectric tensor of the LC.
The free energy density  for such a dielectric has been discussed by  Landau \textit{et al}.\cite{Landau1984} and (at constant charge) it is given by

\begin{eqnarray}
f_\text{E} = {1\over 2} \epsilon_0 \; \pmb{E} \cdot \mbox{\sf\straightepsilon} \cdot \pmb{E}.
\label{DielectricFreeEnergyDensityFromE}
\end{eqnarray}
From $\pmb{\nabla} \cdot \pmb{D}=0$ and $\pmb{\nabla} \times \pmb{E} = 0$, and assuming that the spontaneous polarization is parallel to the molecular dipole vector $\pmb{p}$,
\begin{equation}
\pmb{P}_s = P_0 \; \pmb{p},	
\end{equation}
one can find the electric field as \cite{Perez-Ortiz2011}:
\begin{eqnarray}
E =  - {P_0 \sin{\alpha} \over R \ \epsilon_0 \; \epsilon_{rr} } \int_{R_c}^{R_f}  \left[ {\partial \over \partial R}\left(R \sin{\theta} \right) \right]  \; \text{d}R = - {P_0 \sin{\alpha} \over \epsilon_0 \epsilon_{rr} } \sin{\theta},
\label{ElectricField}
\end{eqnarray}
where $\epsilon_{rr}$ is a component of the dielectric tensor in cylindrical coordinates,
\begin{eqnarray}
\epsilon_{rr} = \epsilon_1 \cos^2{\theta} + \left( \epsilon_2 \cos^2{\alpha} + \epsilon_3 \sin^2{\alpha}  \right) \sin^2{\theta},
\end{eqnarray}
written in terms of the eigenvalues $\epsilon_i$ of that tensor. In Eq.(\ref{ElectricField}) we have assumed that the contribution to the electrostatic free energy in the core region ($R_c$ in Fig \ref{Fig1} b) is so small that we can extend the integration down to zero (i.e., $R_c \rightarrow 0$).

Substitution of Eq.(\ref{ElectricField})  into Eq.(\ref{DielectricFreeEnergyDensityFromE}) yields the final expression for the  electrostatic free-energy  density of the dielectric as a function of the LC orientation:
\begin{eqnarray}
f_\text{E}(\theta,\alpha) = {1 \over 2}  {P_0^2 \sin^2{\alpha} \over \epsilon_0 \epsilon_{rr}} \sin^2{\theta}.
\end{eqnarray}

\subsection{Surface free-energy densities}

We include in our model the energy due to the interfacial tension at the surface of the fiber,
\begin{eqnarray}
f_\text{S} = \pmb {R} \cdot \mbox{\sf\textsigma} \cdot \pmb{R}.
\end{eqnarray}
Here, $\mbox{\sf\textsigma}$ is the biaxial surface-tension tensor:
\begin{eqnarray}
 \mbox{\sf\textsigma} = \sigma_1 \pmb{n} \pmb{n} + \sigma_2 \pmb{m} \pmb{m} + \sigma_3 \pmb{p} \pmb{p},
\end{eqnarray}
so the surface free-energy due to interfacial tension is
\begin{eqnarray}
f_\text{S}\left(\theta(R_f), \alpha \right) = \sigma_1 \cos^2 \theta + \left(   \sigma_2 \cos^2 \alpha + \sigma_3 \sin^2 \alpha   \right) \sin^2 \theta.
\end{eqnarray}

Inhomogeneities in the polarization direction give rise to a bulk free energy with two contributions: the first one is of elastic origin and is due to  packing effects of the mesogens. The second arises from electrostatic energy due to inhomogeneous spontaneous polarization \cite{Bailey2007}:
\begin{eqnarray}
f^\text{bulk}_\text{D}(\theta, \alpha) = c'\left( \pmb{\nabla} \cdot \pmb{p} \right) + c'' \left( \pmb{\nabla} \cdot \pmb{P}_s  \right).
 \end{eqnarray}
 The constants $c'$ and $c''$ correspond to the elastic and electrostatic contributions, respectively. Again, by assuming that  $\pmb{P}_s$ is parallel to $\pmb{p}$, one can integrate this bulk density over the volume of the fiber to obtain a surface density free-energy term \cite{Perez-Ortiz2011}:
\begin{eqnarray}
f_\text{D}\left(\theta(R_f), \alpha \right)= \left( c' +c'' \ P_0  \right) \sin{\alpha} \sin{\theta(R_f)}.
\end{eqnarray}

\section{Monte Carlo simulation}

For the simulation, we write the free energy in non-dimensional form:
\begin{eqnarray}
\mathcal{F}^*= {\mathcal{F}\over 2 \pi L_z \ K}&=& \int_{r_c}^{1} \left[ f^*_\text{N}  +A_\text{L} f^*_\text{L} +A_\text{E} f^*_\text{E} \right] \text{d}r + A_\text{S}f^*_\text{S} +A_\text{D} f^*_\text{D}  ,
\label{IntEneAdi}
\end{eqnarray}
where $L_z$ is the fiber length, $r= R/R_f$, four constants are given by
\begin{eqnarray}
A_\text{L}&=& B R^2_f / K\\
A_\text{E}&=& P_0^2 R_f^2/(\epsilon_0 K)\\
A_\text{S}&=& \sigma_1 R_f /K  \\
A_\text{D}&=& \left( c'+c''P_0   \right) R_f/K ,
\end{eqnarray}
and the free energy densities by
\begin{eqnarray}
f^*_\text{N}(\theta)&=& {1 \over 2} \left( \left( { \text{d} \; \theta\over \text{d} \; r}\right)^2 + {1 \over r^2}  \right) \\
f^*_\text{L}(\theta)&=& {1 \over 2} \left( {L\left(\theta,\alpha\right)\over L_0} -1 \right)^2 \\
f^*_\text{E}(\theta)&=& {1 \over 2} {\sin^2{\alpha} \over \epsilon_{rr}} \sin^2{\theta}  \\
f^*_\text{S}&=& \cos^2{\theta} + \left({\sigma_2 \over \sigma_1} \cos^2{\alpha}  + {\sigma_3 \over \sigma_1} \sin^2{\alpha}   \right) \sin^2{\theta}\\
f^*_\text{D}&=& \sin{\alpha} \sin{\theta}.
\end{eqnarray}

By assuming cylindrical symmetry, we solve for $\theta(r)$ on a one-dimensional mesh with $N= 61$ or $N=81$ nodes along the radial coordinate. The mesh starts one coherence length away from the center and ends at the surface of the fiber. Since we anticipate that $\theta(r)$ varies slowly away from the fiber surface, we place half of the nodes equidistantly between the surface and 15 coherence lengths below it. The other half is distributed equidistantly in the last 15 coherence lengths. We then estimate numerically the integral in (\ref{IntEneAdi}) by the sum
\begin{eqnarray}
\mathcal{F}^* \approx  \sum_{i=1}^N \left( f^*_\text{N}(\theta_i)  +A_\text{L} f^*_\text{L}(\theta_i)  +A_\text{E} f^*_\text{E} \right) r_i \Delta r_i + A_S f^*_S + A_D f^*_D,
\end{eqnarray}
where $\theta_i = \theta(r_i)$ and $r_i$  is the location of the $i$-th node.

The simulations were started by setting all $\theta_i$ to a single value, as described below.  We then iterate Metropolis Monte Carlo steps as follows: from the old configuration, $\theta^o$, we select with uniform probability the value at the $j$-th node, $\theta_j^n$. An update is then proposed to a new value:
\begin{equation}
\theta_j^n = \theta_j^o + \Delta \left( \xi - 0.5 \right),
\end{equation}
where $\xi$ is a random number (distributed uniformly between zero and one) and  $\Delta$ is a parameter that controls the maximum increment to $\theta_j^o$.  The new value is accepted with probability\cite{Metropolis1953}
\begin{equation}
P_\text{acc} = \text{min} \left(1, e^{-\beta^* \Delta \mathcal{F^*}}   \right),
\end{equation}
where $\Delta  \mathcal{F}^*$ is the difference of free energies between the new and original configurations:
\begin{eqnarray}
\Delta \mathcal{F}^* = \mathcal{F}^*\left[ \theta^n \right] - \mathcal{F}^*\left[ \theta^o \right],
\end{eqnarray}
For each simulation, constant $\Delta$ was chosen so that acceptance of the proposed configuration was between 20\% and  50\% \cite{Frenkel2001}.  By changing the value of Metropolis parameter  $\beta^*$, we implemented a simulated annealing method \cite{Kirkpatrick1983,Londono-Hurtado2015}. In a typical simulation, we iterate the Metropolis MC steps, and every 10 million of them we anneal the system.  The results were analyzed with Mathematica 8.0 \cite{Wolfram2010}.
\section{Structural transition from an homogeneous-bulk estimate}

Prior to our simulations, we studied the behavior of the non-dimensional bulk free energy as a function of the constant value of a uniform orientation field, $\theta(r)= \theta_\text{bulk}$, for two cases:  first, by changing the parameter $\Theta_0$ of flat-layer orientation at fixed external radius $R_f$, and then by changing $R_f$ at fixed $\Theta_0$.

\begin{figure}[ht]
\begin{center}
\includegraphics[width=8cm,height=!]{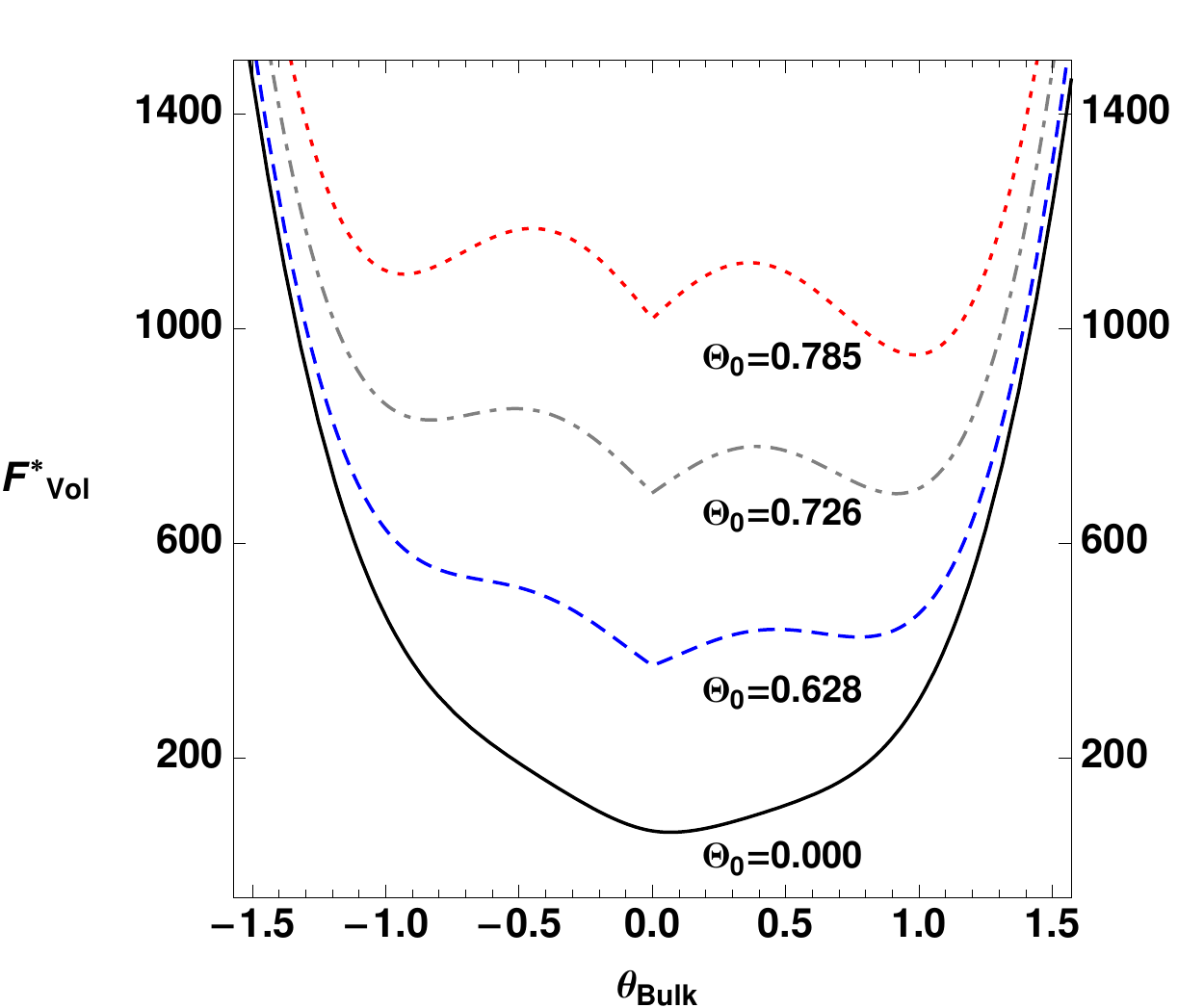}
\end{center}
\caption{The volumetric free energy $F^*_\text{Vol}$ of a liquid crystal fiber as a function of the bulk orientation $\theta_\text{bulk}$ displays a first-order transition, depending on the  angle $\Theta_0$ corresponding to the flat-layer orientation of the LC. For $\Theta_0$=0 (solid line) the free energy has a single minimum at $\theta_\text{bulk}$=0 but for larger values other minima appear. At  $\Theta_0=0.628$ (dashed line) the origin is a stable minimum, but there is a first-order transition at $\Theta_0 = 0.726$ (dot-dashed line). At $\Theta_0 = 0.785$ (dotted line), the state with $\theta_\text{bulk} > 0$ is the stable one.
%
 For clarity, the curves have been shifted up by 250, 500 and 750 for the three larger values of $\Theta_0$, respectively.}
\label{EneTheAnalytic}
\end{figure}

Figure~\ref{EneTheAnalytic} shows the bulk free energy for several values of the flat-layer orientation $\Theta_0 = 0.000,\; 0.628,\; 0.726,\; 0.785$, for constant external radius  $R_f= 1.0 \; \mu$m. For small values of $\Theta_0$, there exists only a minimum at $\theta_\text{bulk}=0$. This minimum is non-differentiable, due to the absolute value in the layer compression term. Since the derivative is not defined at the origin, it cannot be found by setting it to zero. As one increases the value of $\Theta_0$, the bulk free energy acquires two additional minima. They are metastable with respect to the non-differentiable minimum until $\Theta_0$ reaches the transition value 0.726. For larger values of $\Theta_0$, the stable minimum is the one with $\theta_\text{bulk} > 0$. Since the change in the value of the stable minimum at the transition is discontinuous, this model predicts a first-order transition.
\

\begin{figure}[ht]
\begin{center}
\includegraphics[width=8cm,height=!]{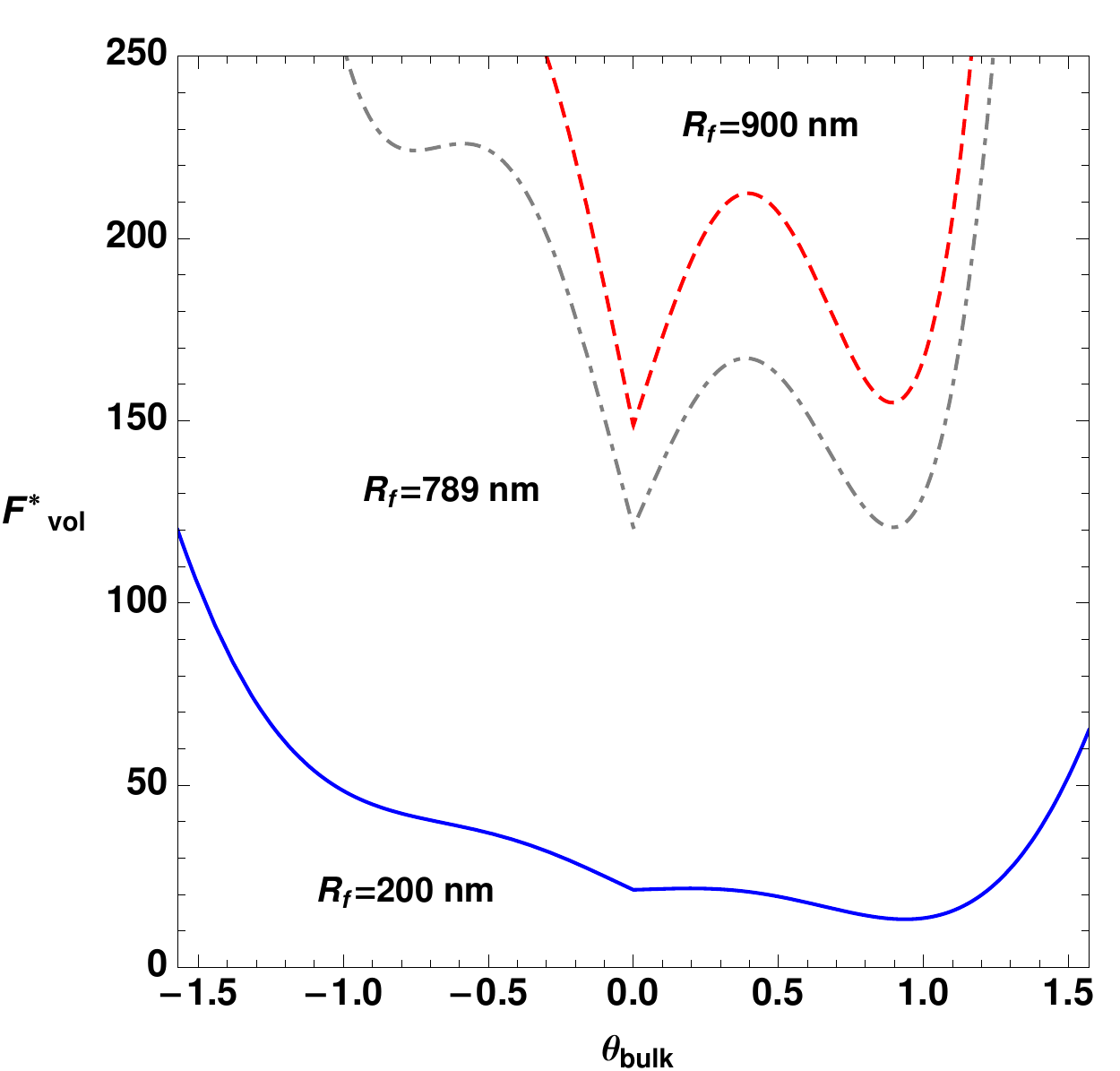}
\end{center}
\caption{ The  volumetric free energy $\mathcal{F}^*_\text{vol.}$ of the fiber, as a function of the bulk orientation $\theta_\text{bulk}$, shows another first-order transition when one changes the external radius $R_f$ while keeping $\Theta_0=0.691$ fixed. For small radii, such as $R_\text{f}=200$~nm (solid line) the stable minimum occurs at positive $\theta_\text{bulk}$; for larger values (e.g. $R_\text{f}=900$ nm, dashed line) the stable minimum is at the origin. The transition is located at $R_f= 789$ (dot-dashed line).
}
\label{EneRadAnalytic}
\end{figure}

There is also a first-order transition that depends on the external radius of the fiber; in our simulations, we are able to set specific values of the fiber radius, even if this may be difficult to achieve experimentally. In order to show the transition, Fig.~\ref{EneRadAnalytic} displays the behavior of the bulk free energy (at fixed flat-layer orientation $\Theta_0= 0.691$) for the external radii $R_f= 170, \; 672, \; 900$ nm.  As before, we observe three minima and the bulk free energy is non-differentiable at $\theta_\text{bulk}=0$. When the radius is small, the stable minimum is that for  $\theta_\text{bulk} > 0$, but there is a transition when $R_f= $ 789~nm and, for larger values, the stable minimum is that with $\theta_\text{bulk} = 0$.

\

\section{Structural transition from Monte Carlo simulation }

In our Monte Carlo simulations we allow for non-uniform fields $\theta(r)$ and take into account both the bulk and surface terms in the free energy. We set the material parameters to the values given in Table~\ref{Tabla1}, taken from Bailey et al.\cite{Bailey2007} and P\'erez-Ortiz et al.\cite{Perez-Ortiz2011}.

\begin{table}[htb]
\begin{center}
\begin{tabular}{c c l }
\hline \hline
\ \ \ \ Symbol \ \ \ \  & \ \ \ \ \ \ \ \ \ \ \ \ \ \ \ \ \ \ \ \ \ \ \ \ \ \ \ Value  \ \ \ \ \ \ \ \ \ \ \ \ \ \ \ \ \ \ \ \ \ \ \ \ \ \ \  &  Parameter  \ \ \ \ \ \ \ \ \ \ \ \ \ \ \ \ \ \ \ \ \ \ \ \  \\ \hline
$\Theta_0$		& $0\le\Theta_0 \le \pi/4$			& Flat-layer orientation						\\
$\alpha_0$ 		& $0$						& Flat-layer polarization angle				\\
$\alpha$			& $-\pi/2$						& Polarization angle 						\\
$R_c$			& $5.0$ nm					& Core radius								\\
$K$				& $10^{-11}$ N					& Elastic constant							\\
$B$				& $10^{5}$ Pa					& Layer compression modulus					\\
$L_1$			& $1.5$ nm					& Medium axis of mesogens					\\
$L_2$			& $0.5$ nm					& Short axis of mesogens						\\
$L_3$			& $5.0$ nm					& Long axis of mesogens 						\\
$P_0$			& $10^{-3}$  C m$^{-2}$			& Spontaneous polarization					\\
$c'$				& $0.017$  N/m					& Elastic Inh. of dipole direction constant			\\
$c''$				& $11.2$ N m C$^{-1}$			& Electric Inh. of dipole direction constant			\\
$\epsilon_1$		& $7$						& Dielectric constant in direction $\pmb{n}$		\\
$\epsilon_2$		& $10$						& Dielectric constant in direction $\pmb{m}$		\\
$\epsilon_3$		& $12$						& Dielectric constant in direction $\pmb{p}$		\\
$\sigma_1$		& $0.026$  N m$^{-1}$			& Surface tension in direction $\pmb{n}$			\\
$\sigma_2$		& $0.024$  N m$^{-1}$			& Surface tension in direction $\pmb{m}$			\\
$\sigma_3$		& $0.025$  N m$^{-1}$			& Surface tension in direction $\pmb{p}$			\\
 \hline \hline
\end{tabular}
\end{center}
\caption{The parameters for the liquid crystal used in our simulations are taken from the works of Bailey et al.\cite{Bailey2007} and P\'erez-Ortiz et al.\cite{Perez-Ortiz2011}.}
\label{Tabla1}
\end{table}

\subsection{Structural transition with the flat-layer orientation $\Theta_0$}

In order to test for the first transition, we run simulations at constant radius $R_f=1.0 \mu$m and vary $\Theta_0$.  At $\Theta_0= 0.785$, we had originally expected a stable state with $\theta(r) >0$ and a metastable state with $\theta(r) =0$. Therefore, we initialized the system to the uniform field $\theta(r)=0$ and tracked the total free energy of the system.

\begin{figure}[ht]
\begin{center}
\includegraphics[width=8cm,height=!]{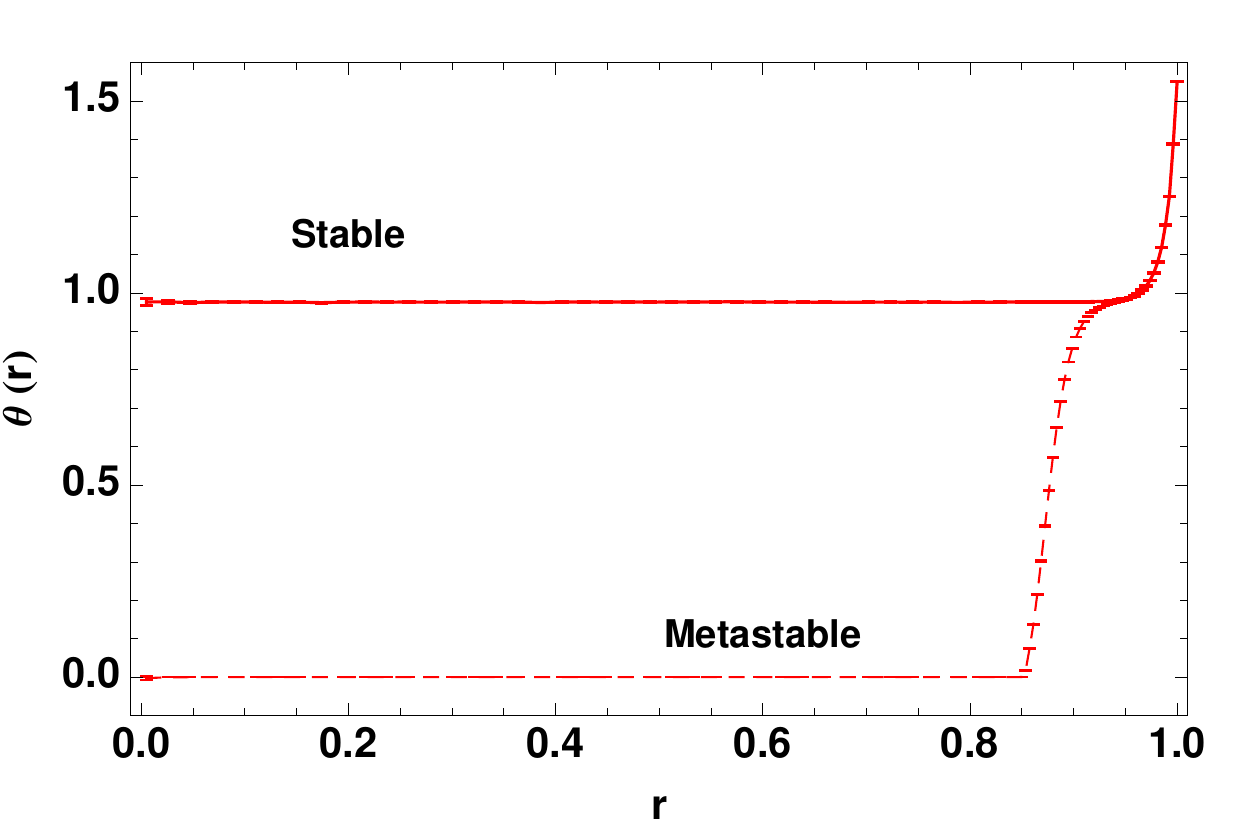}
\includegraphics[width=8cm,height=!]{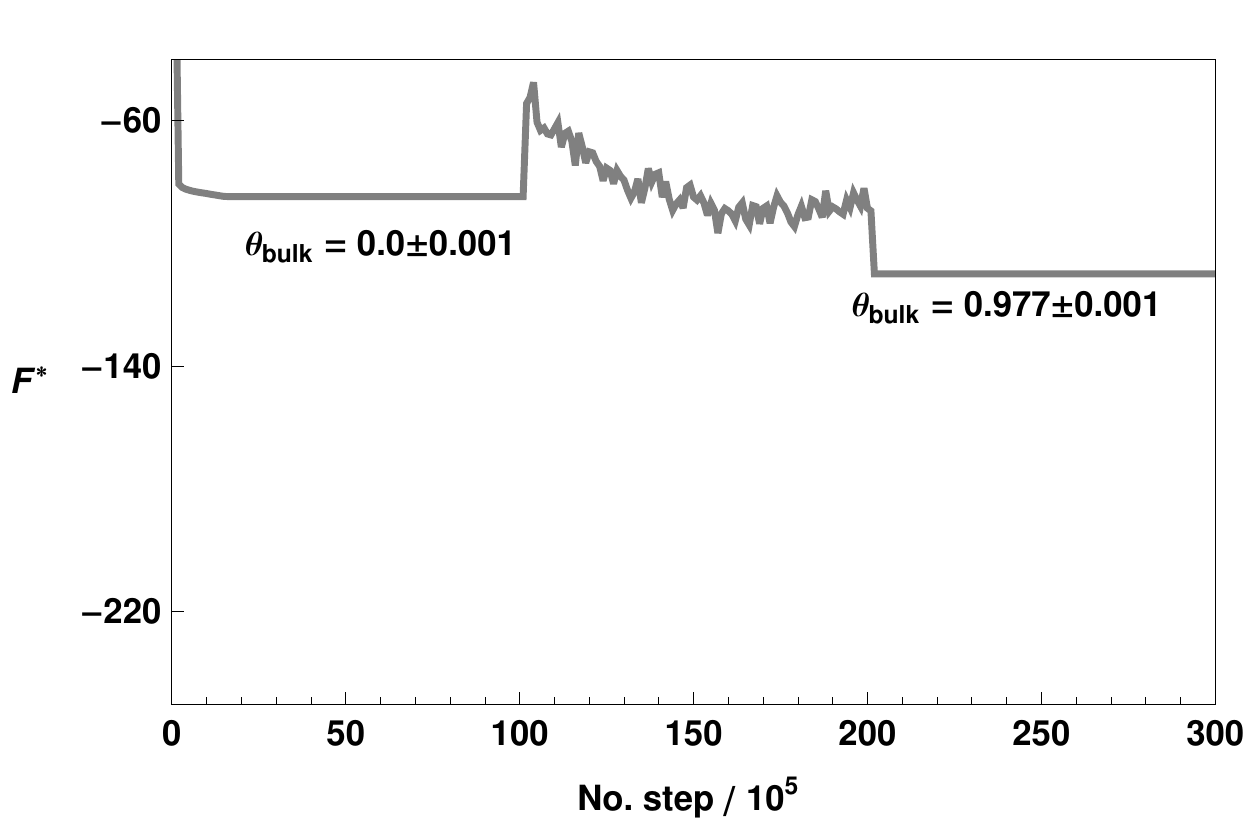}
\includegraphics[width=8cm,height=!]{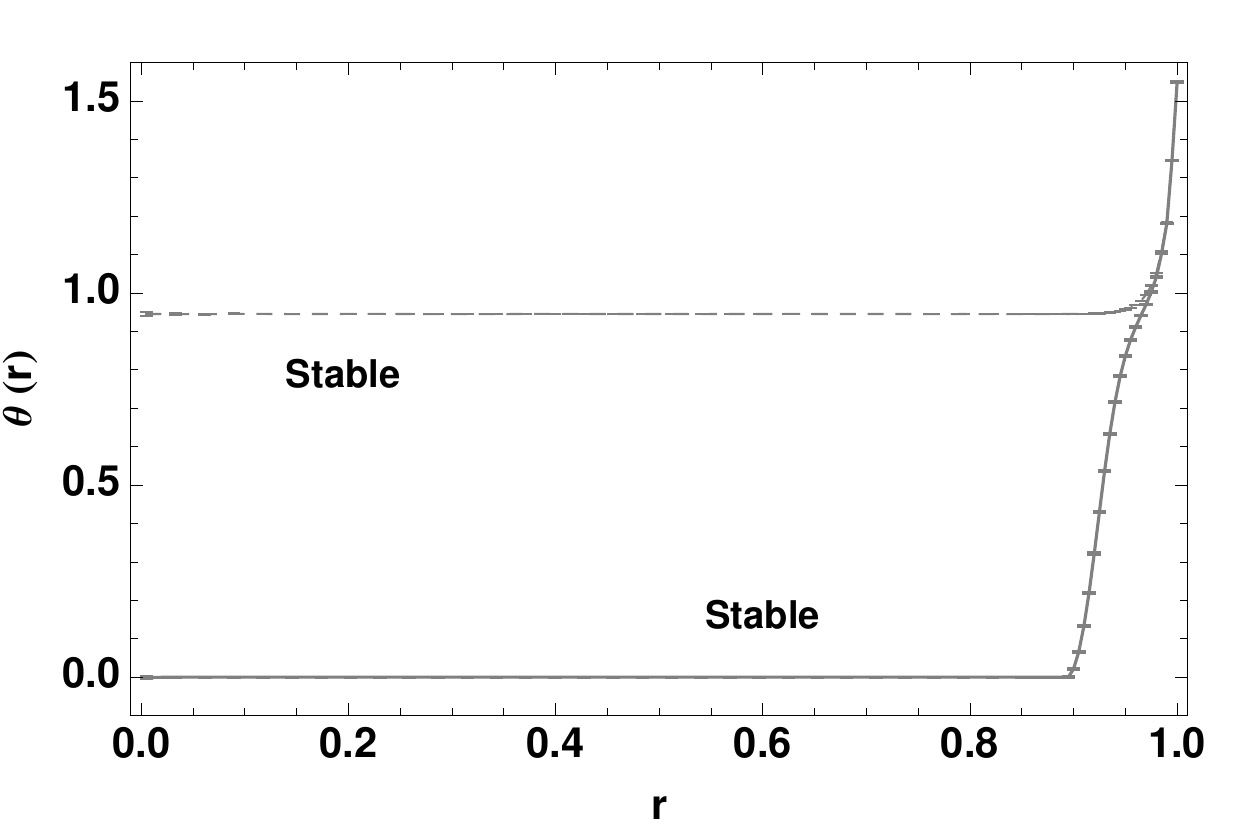}
\includegraphics[width=8cm,height=!]{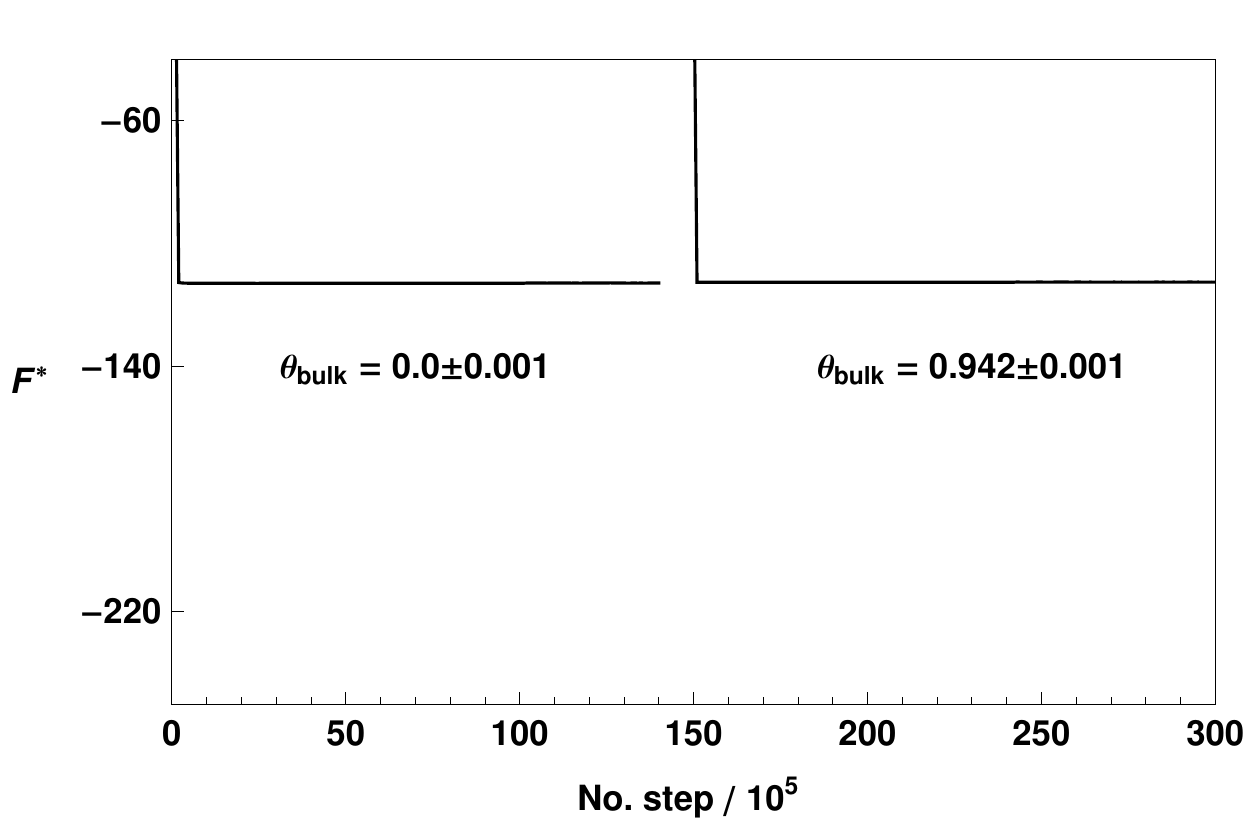}
\includegraphics[width=8cm,height=!]{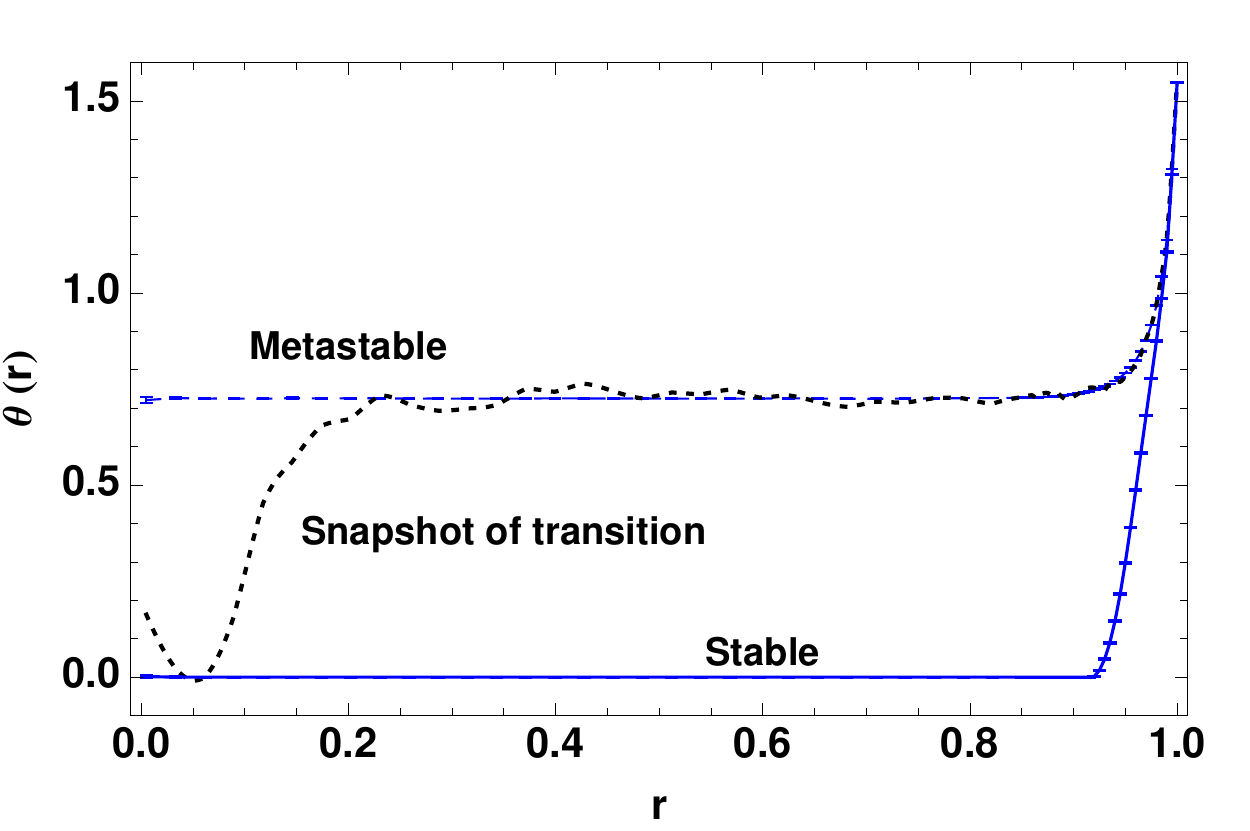}
\includegraphics[width=8cm,height=!]{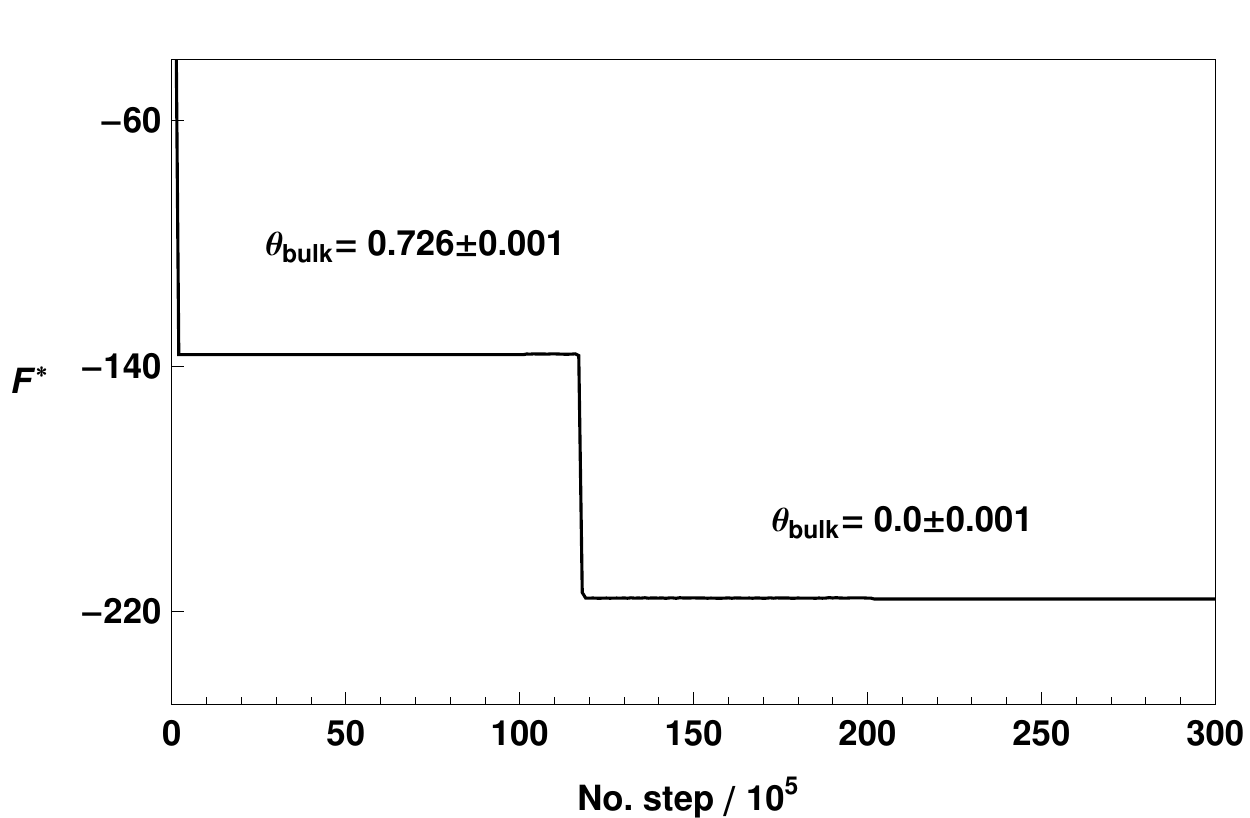}
\end{center}
\caption{Monte Carlo simulations of the LC fiber show a first-order transition with the flat-layer orientation angle $\Theta_0$ between orientational structures with zero and positive $\theta_\text{bulk}$: (a) at $\Theta_0 = 0.785$, the configuration with $\theta_\text{bulk}=0$ is metastable and that with $\theta_\text{bulk} > 0$  is stable (see b); (c) at $\Theta_0 = 0.760$ both states are stable, since they have the same free energy (see d); and (e) at $\Theta_0 = 0.628$, the first state has become stable and the second metastable (see f). The radius of the fiber was kept fixed at $R_f = 1 \mu$m. In (a) and (c) the boundary layer of the metastable states display a kink as a stable value of $\theta_\text{bulk}$ is crossed, but no kink is visible in (e) since no stable value is crossed.
%
%
%
%
 }
\label{EneThe25}
\end{figure}

We observed that indeed the system adopts one of two  configurations: the first one tends to $\theta_\text{bulk} = 0.000 \pm 0.001$ as $r \rightarrow 0$, the other tends to $\theta_\text{bulk}=0.977 \pm 0.001$ (see  Fig.~\ref{EneThe25}a). Both configurations display boundary layers. The first configuration has higher free energy than the second, as shown in Fig.~\ref{EneThe25}b, and therefore it is confirmed to be metastable with respect to the latter.

At first, since the system was initialized to $\theta(r)=0$ and $\beta^*= 2\times 10^3$, the system adopted the metastable configuration. The stable configuration was obtained through an annealing procedure, by temporarily decreasing the Metropolis parameter ($\beta^*=2$) until the system abandoned the metastable state, and then resetting it to its original value. The long-dashed and solid lines in Fig.~\ref{EneThe25}b are averages over the metastable and stable states, respectively, over the corresponding plateaus displayed in Fig.~\ref{EneThe25}b.

There is an additional indication of the presence of the stable state in the bulk free energy: when the system is in the metastable state with $\theta_\text{bulk} = 0.000 \pm 0.001$, the boundary layer that goes continuously from zero to $\pi/2$ shows a kink as it passes through the stable-mimimum value $0.977$. This is because the mesh nodes with values close to the stable minimum are less likely to change when subjected to the Metropolis criterion.

\

We next simulated the case with $\Theta_0 =0.628$, where a stable state was expected with $\theta(r) =0 $ and a metastable state  with  $\theta(r) > 0 $. Initializing the system to the uniform field $\theta(r)=1.0$ and $\beta^*=5 \times 10^3$, we observed again two different equilibrium configurations: the first one tends to  $\theta_\text{bulk}=0.000 \pm 0.001$  away from the surface, the second tends to  $\theta_\text{bulk} = 0.726 \pm 0.001$ (see  Fig.~\ref{EneThe25}e). These values are close to the minima estimated solely from the bulk free energy: 0 and $0.776$, respectively. Despite the presence of boundary layers, $\theta_\text{bulk}$ in the simulation differs less than 7\% from the value predicted from the homogeneous-bulk approximation. This indicates that the surface effects are very short ranged and they do not affect the fiber interior too much.

Since the simulation was initialized to $\theta(r)=1.0$, the system adopted the metastable configuration first. Again, the stable configuration was obtained with an annealing procedure, by temporarily decreasing $\beta^*=1.5 \times 10^2$ until the system abandoned the metastable state, and then resetting it to its original value. In Fig.~\ref{EneThe25}e we show a snapshot of the orientation field as it transits from the metastable state to the stable one; since the MC simulation is based on local changes to the values of the orientation field, we observe that some nodes in the simulation mesh overcome the energy barrier and then the transition propagates as a wave.
%

We located the transition point in the MC simulations by requiring that the free energies of the coexisting minima be equal. We found the value $\Theta_0=0.760$, which is near to  the homogenous-bulk estimate $\Theta_0=0.726$. Figure~\ref{EneThe25}c shows the two coexisting states at $\Theta_0=0.760$ and $\beta^*=5 \times 10^3$: the first one starts close to zero and then develops a boundary layer. The second state was found by initializing with the uniform field $\theta(r)=1.0$; it tends to the bulk value $0.942 \pm 0.001$ away from its boundary layer. Only the boundary layer of the first state shows a kink, since it is the only one that has to pass through the other minimum to reach its value at the surface.

%




\subsection{Structural transition with the external radius $R_f$}

\begin{figure}[ht]
\begin{center}
\includegraphics[width=8cm,height=!]{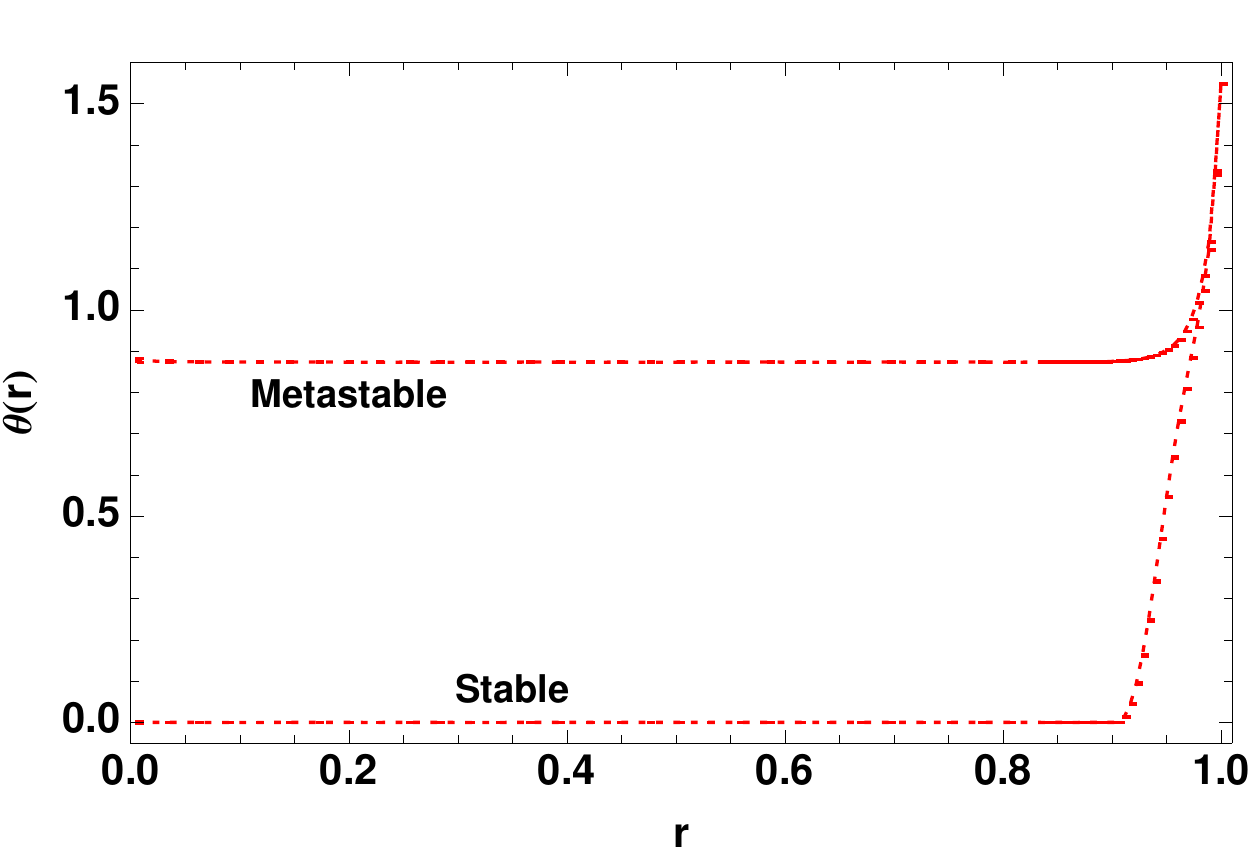}
\includegraphics[width=8cm,height=!]{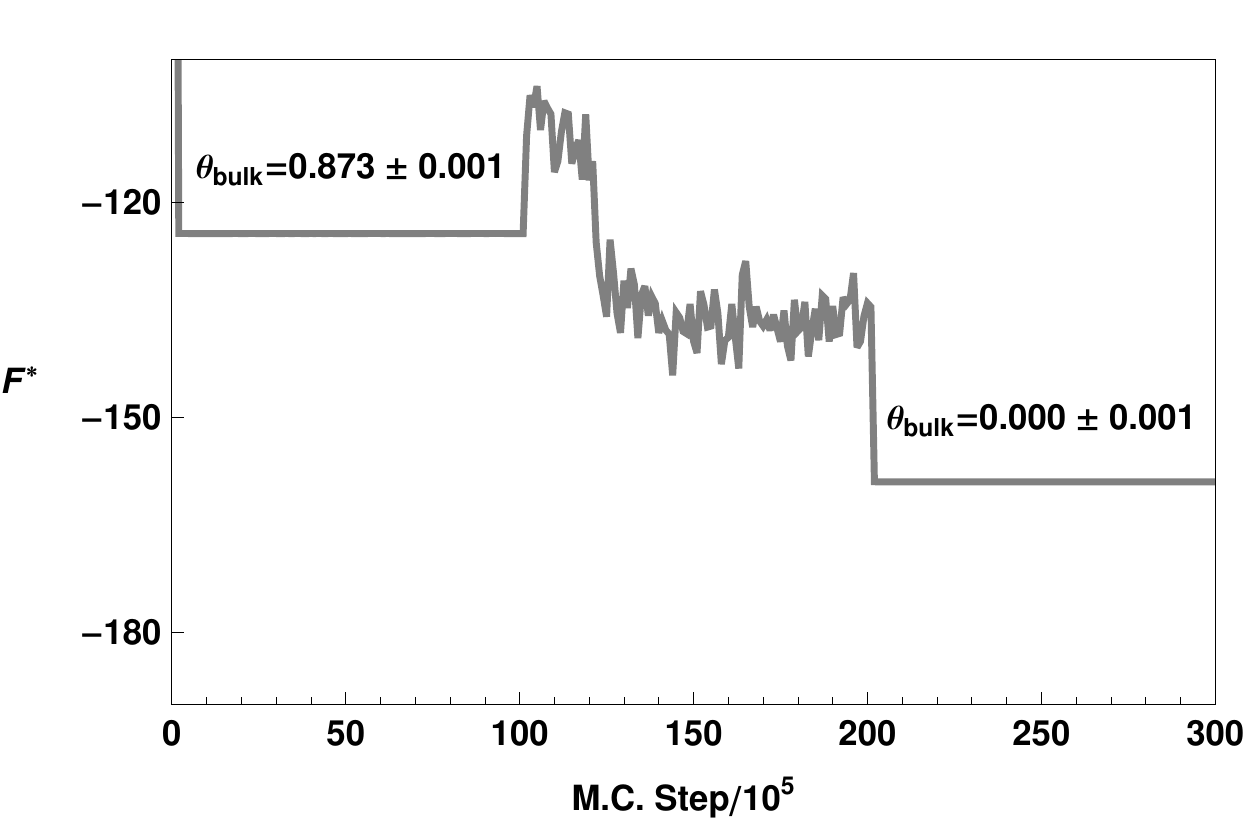}
\includegraphics[width=8cm,height=!]{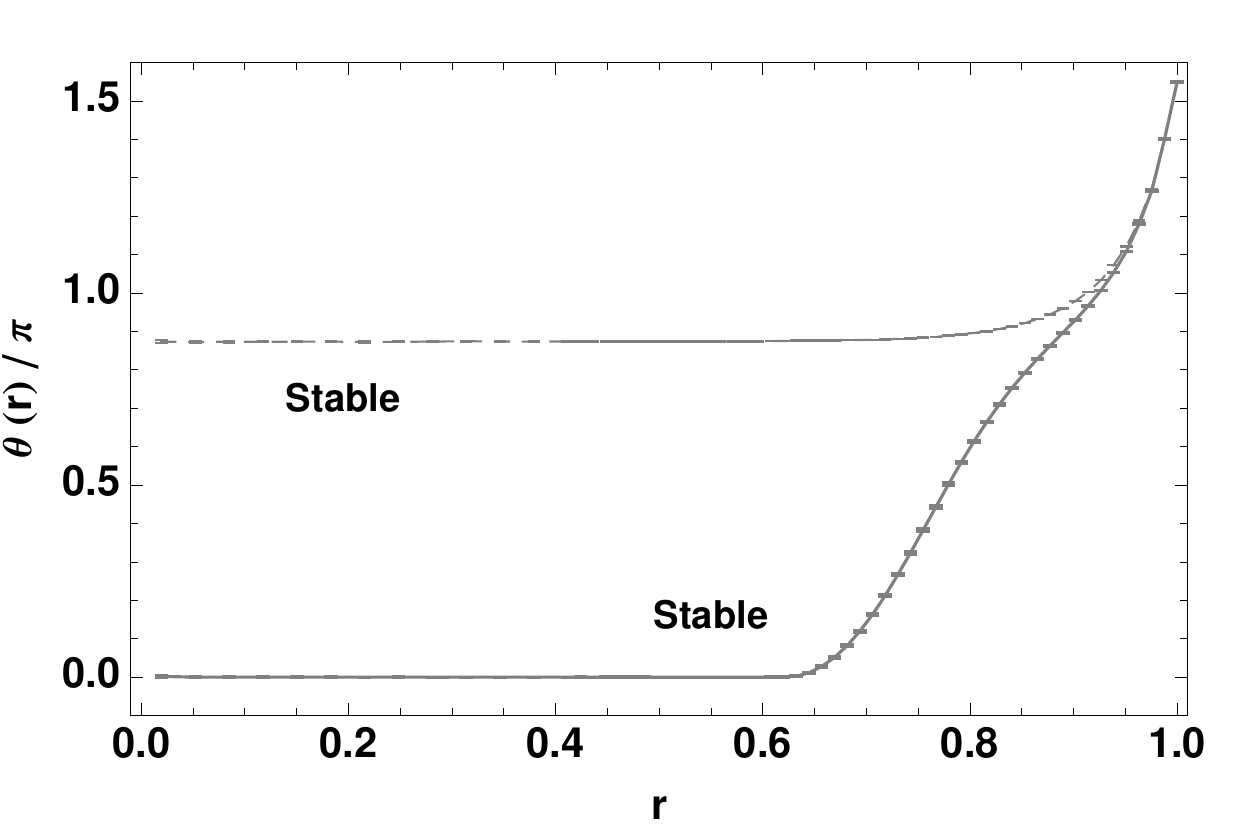}
\includegraphics[width=8cm,height=!]{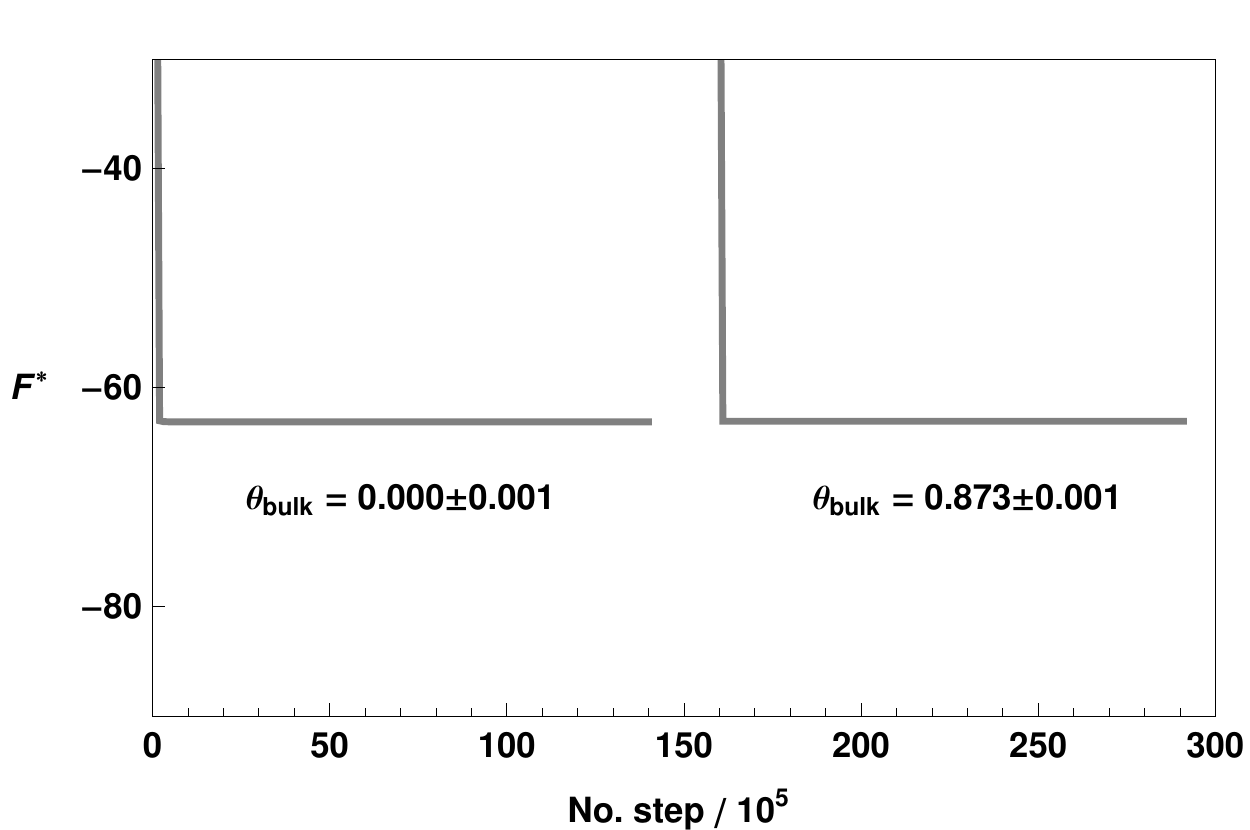}
\includegraphics[width=8cm,height=!]{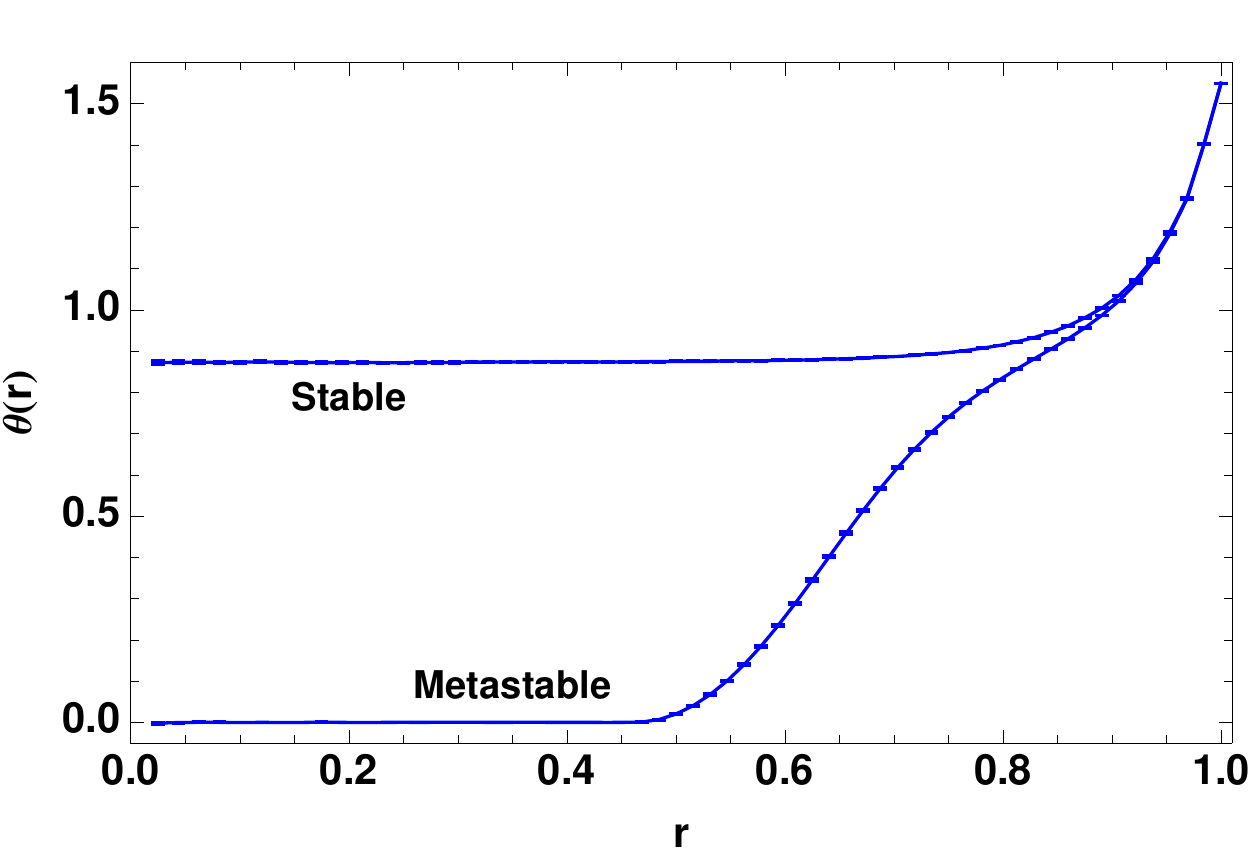}
\includegraphics[width=8cm,height=!]{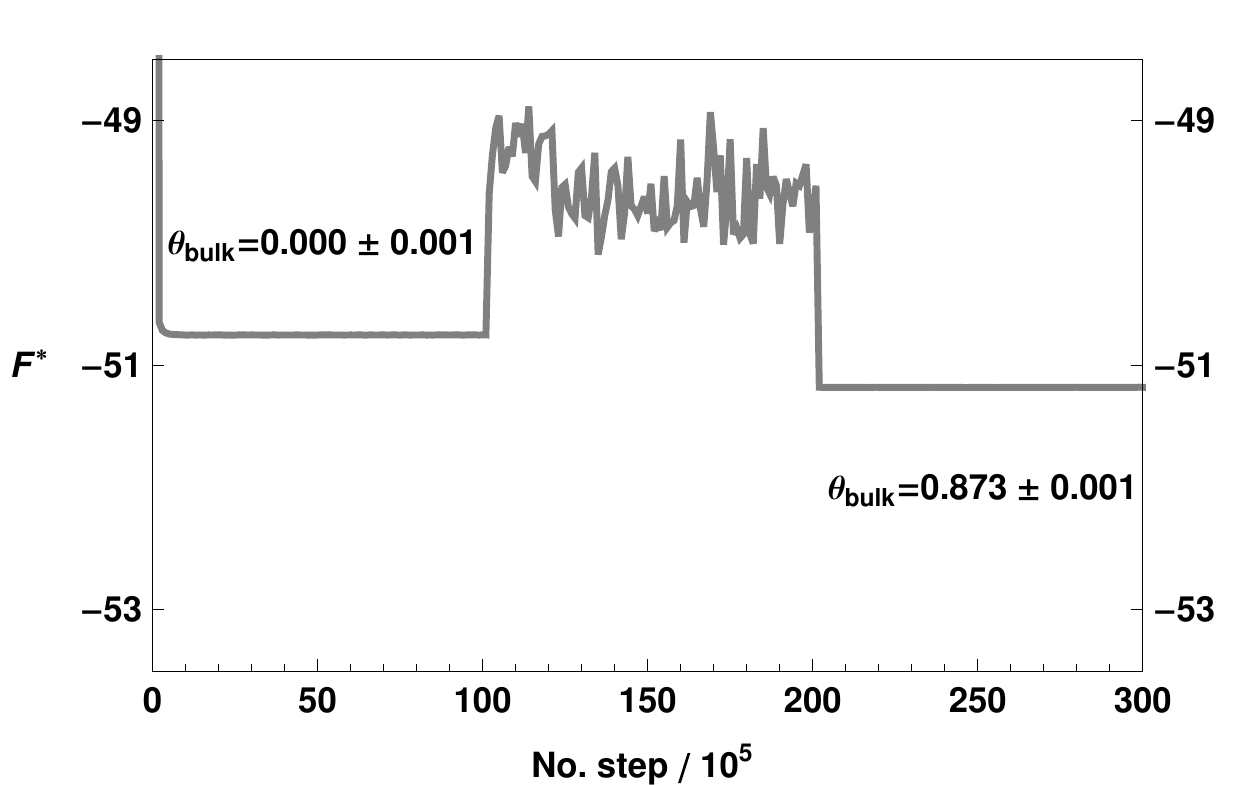}
\end{center}
\caption{
Monte Carlo simulations of the LC fiber also show a first-order transition with the fiber radius $R_f$ between orientational structures with zero and positive $\theta_\text{bulk}$: (a) at $R_f = 900$~nm, the configuration with $\theta_\text{bulk}=0$ is stable and that with $\theta_\text{bulk} > 0$  is metastable (see b); (b) at $R_f = 0.789$~nm both states have the same free energy; and (c) at $R_f = 200$~nm, the first state has become metastable and the second stable (see f). The flat-layer orientation angle $\Theta_0 = 0.691$ was kept fixed. Again, in (b) and (c) the boundary layers display a kink as a stable value of $\theta_\text{bulk}$ is crossed, but no kink is visible in (a) since no stable value is crossed.
}
\label{SimRad020}
\end{figure}

Figure~\ref{SimRad020} shows the configurations obtained by Monte Carlo simulations (at constant flat-layer orientation $\Theta_0 = 0.691$) confirming the presence of stable and metastable states on both sides of the first-order transition, as a function of the external radius of the fiber.

For the larger radius $R_f = 900$~nm, we started the simulation with a uniform field $\theta(r) = 1.0$ and $\beta^*=5 \times 10^3$.  At first, the system stayed in a metastable state with  $\theta_\text{bulk}=0.873 \pm 0.001$  but after annealing (with $\beta^*=2$) it reached the stable configuration with $\theta_\text{bulk}= 0.000 \pm 0.001$.


For the smallest radius, 200 nm, we initialized to the uniform field $\theta(r)=0.0$ and $\beta^*=1 \times 10^4$ and found a state with $\theta_\text{bulk} = 0$. Then, by temporarily decreasing $\beta^*$ to $2\times 10^2$ in the annealing procedure, the system changed to the stable state with $\theta_\text{bulk} = 0.873 \pm 0.001$. Analysis of the free energies of both states shows that the first one is only metastable.

We located the transition in the simulations at $R_f=255$ nm, which is far from the value $R_f=789$ nm obtained by neglecting the surface contribution. This is because the surface contribution becomes increasingly important for small radii, as is the case at the transition. The two coexisting states were found by initializing to two different conditions: $\theta(r)=0.0$ and $\beta^*= 1\times 10^4$ for the state with $\theta_\text{bulk}=0.000 \pm 0.001$;  $\theta(r)=1.0$ and $\beta^*= 1\times 10^4$ for the configuration with $\theta_\text{bulk}=0.873 \pm 0.001$.



\begin{figure}[ht]
\begin{center}
\includegraphics[width=8cm,height=!]{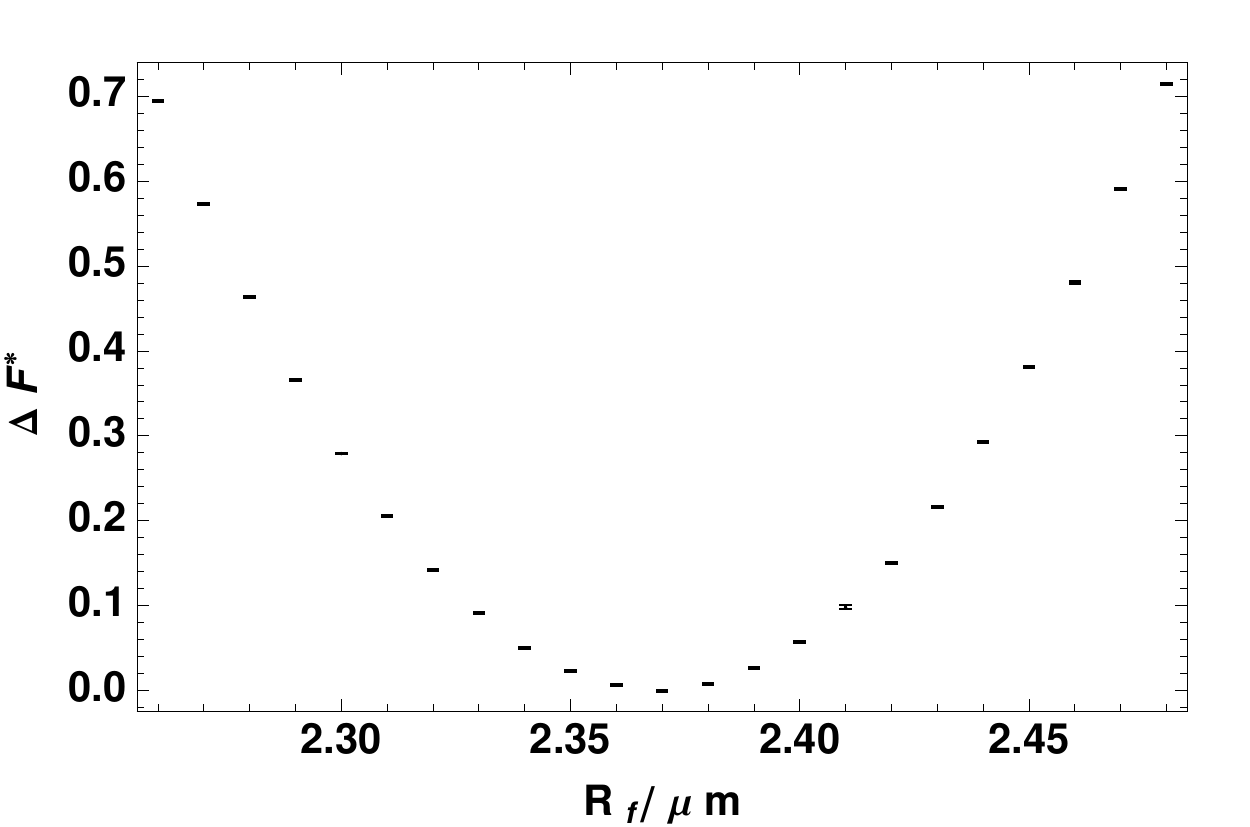}
\end{center}
\caption{The free energy of the stable equilibrium states from our simulations, as a function of the external radius $R_f$, show a minimum at $R_f~=~2.37~\mu$m, when the physical parameters of the liquid crystal are assumed within their experimental ranges \cite{Bailey2007,Perez-Ortiz2011} (see Table \ref{Tabla1}). Thus, we can recover the experimentally observed size of the LC fibers. Error bars indicate the standard error of the  free energy in each simulation.}
\label{Fig7}
\end{figure}

Finally, we found the equilibrium radius that corresponds to a fixed value of the flat-layer orientation $\Theta_0$, by tracking the stable states of a succession of simulations with increasing radii, using the material parameters previously used by P\'erez-Ortiz et al.\cite{Perez-Ortiz2011} and Bailey et al.\cite{Bailey2007} Figure~\ref{Fig7} shows that the total free energy of the stable states displays a definite minimum as a function of the external radius $R_f$, corresponding to an equilibrium radius of 2.37~$\mu$m. This value is within the the range of reported experimental observations \cite{Jakli2003,Chen2013}.

By repeating the sequence of simulations with a different value of the parameter $c'$, we found that the equilibrium radius is particularly sensitive to the surface free energy due to inhomogeneities in the polarization. A doubling of $c'$  yields a tenfold increment of the equilibrium radius: specifically, going from $c'= 0.017$ N/m to 0.040 N/m results in a change of $R_f$ from 2.37 $\mu$m to 22 $\mu$m. From these observations, we predict that changes in the surface tension may also affect greatly the equilibrium radius.

\

\section{Conclusion}

In this work we presented field-theory Monte Carlo simulations using a free energy model with both bulk and surface contributions. We confirmed that the equilibrium configurations display a boundary layer of about 15 coherence lengths as well as a largely constant plateau, as predicted by  {P\'erez-Ortiz} \textit{et al.}\cite{Perez-Ortiz2011}.

We analyzed the bulk free energy and found that it predicts a first-order transition for the director orientation, both as a function of the flat-layer angle
$\Theta_0$ and the fiber radius $R_f$. We confirmed this with our Monte Carlo method and identified a perturbing effect of the stable minima on the boundary layers of coexisting or metastable configurations.
Both stable and metastable equilibrium configurations can be used as inputs for models of the propagation of light along the LC fibers.

A straightforward generalization of our Monte Carlo methodology is to introduce a second orientation field, $\alpha(r)$ so that the director no longer is confined to planes perpendicular to the fiber axis. This would enable one to address escaped configurations \cite{deGennes1993} that lack central defects.

Regarding defects, our present model cannot describe the behavior of the LC at their core because of the divergence of Frank elasticity. However, it would be possible to include the defects by expressing the bulk free energy in terms of the tensor order parameter ({\bf Q}), as in the model proposed by Mukherjee \cite{Mukherjee2004} for bent-core smectics in the bulk and proposing the corresponding surface energy terms.

\section*{Acknowledgments}
N. Atzin acknowledges the support from Universidad Aut\'{o}noma Metropolitana for a Ph.D. scholarship.

%

\end{document}